\documentclass[twocolumn,preprintnumbers,amsmath,amssymb]{revtex4}
\usepackage{dcolumn}
\usepackage{bm}
\usepackage{graphicx,subfigure,xcolor}
\usepackage{braket}
\usepackage[normalem]{ulem}
\usepackage{balance}
\usepackage{comment}
\usepackage{xcolor}
\usepackage{amsmath}
\usepackage{amssymb}
\usepackage{eucal}
\usepackage{mathrsfs}
\usepackage{amsthm}
\usepackage{epstopdf}
\usepackage{textcomp}
\usepackage{braket}
\usepackage{pdfpages}
\usepackage[utf8]{inputenc}
\usepackage[T1]{fontenc}
\usepackage{soul}
\usepackage{hyperref}
\hypersetup{colorlinks,
linkcolor=blue,
filecolor=green,
urlcolor=blue,
citecolor=blue}

\newcommand{\wo}{\omega_0}
\newcommand{\wk}{{\omega_k}}

\def\bk{{\bf k}}

\def\br{{\mathbf r}}

\def\bR{{\mathbf R}}

\def\bfm{{\bf f}}

\newcommand{\bmu}{\boldsymbol\mu}

\newcommand{\skj}{\sum_{{\bf k}j}}

\newcommand{\akjd}{a^{\dag}_{{\bf k}j}}
\newcommand{\akj}{a_{{\bf k}j}}

\newcommand{\ekj}{\hat{{\bf e}}_{{\bf k}j}}

\newcommand{\tp}{t^\prime}
\newcommand{\tpp}{t^{\prime\prime}}

\begin{document}

\title{Collective spontaneous emission of two entangled atoms near an oscillating mirror}

\author{Marta Reina$^{1}$\footnote{marta.reina@institutoptique.fr}}
\author{Michelangelo Domina$^{2}$\footnote{dominam@tcd.ie}}
\author{Alessandro Ferreri$^{3}$\footnote{alessandro.ferreri@uni-paderborn.de}}
\author{Giuseppe Fiscelli$^{1}$\footnote{giuseppe.fiscelli@unipa.it}}
\author{Antonio Noto$^{1}$\footnote{antonio.noto@unipa.it}}
\author{Roberto Passante$^{1,4}$\footnote{roberto.passante@unipa.it}}
\author{Lucia Rizzuto$^{1,4}$\footnote{lucia.rizzuto@unipa.it}}

\affiliation{$1$ Dipartimento di Fisica e Chimica - Emilio Segr\`{e}, Universit\`{a} degli Studi di Palermo, Via Archirafi 36, I-90123 Palermo, Italy}
\affiliation{$2$ School of Physics and CRANN Institute, Trinity College, Dublin 2, Ireland}
\affiliation{$3$ Department of Physics, Paderborn University, Warburger Strasse 100, D-33098 Paderborn, Germany}
\affiliation{$4$ INFN, Laboratori Nazionali del Sud, I-95123 Catania, Italy}

\begin{abstract}
We consider the cooperative spontaneous emission of a system of two identical atoms, interacting with the electromagnetic field in the vacuum state and in the presence of an oscillating mirror.
We assume that the two atoms, one in the ground state and the other in the excited state, are prepared in a correlated (symmetric or antisymmetric) {\em Bell}-type state. We also suppose that the perfectly reflecting plate oscillates adiabatically, with the field modes satisfying the boundary conditions at the mirror surface at any given instant, so that the time-dependence of the interaction Hamiltonian is entirely enclosed in the instantaneous atoms-wall distance.
Using time-dependent perturbation theory, we investigate the spectrum of the radiation emitted by the two-atom system, showing how the oscillation of the boundary modifies the features of the emitted spectrum, which exhibits two lateral peaks not present in the case of a static boundary. We also evaluate the transition rate to the collective ground state of the two-atom system in both cases of the superradiant (symmetric) and subradiant (antisymmetric) state. We show that it is modulated in time, and that the presence of the oscillating mirror can enhance or inhibit the decay rate compared to the case of atoms in vacuum space or near a static boundary. Our results thus suggest that a dynamical (i.e. time-modulated) environment can give new possibilities to control and manipulate radiative processes of atoms or molecules nearby, such as the cooperative decay, and strongly indicate a similar possibility for other radiative processes, for example the resonance interaction and the energy transfer between atoms or molecules.
\end{abstract}

\maketitle

\section{\label{sec:1}Introduction}

Quantum electrodynamics predicts that an excited atom, interacting with the quantum electromagnetic field in the vacuum state, spontaneously decays to its ground state by emitting a photon. The emission probability for unit time is found to be
\begin{eqnarray}
\label{eq:1}
A=\frac 4 3\frac{\omega_{eg}^3 \mid\bmu^{eg}\mid^2}{\hbar c^3}\, ,
\end{eqnarray}
where $\bmu^{eg}$ is the matrix element of the atomic dipole moment operator between the atomic excited and ground states, and $\omega_{eg}$ is the transition frequency between the two atomic levels \cite{dirac}.  This result can be generalized to the case of $N$ atoms
incoherently coupled to the quantum electromagnetic field: in this case, the $N$ atoms decay independently, and the intensity of the emitted radiation is proportional to $N$.
Dicke in $1954$ \cite{dicke54} showed that this conclusion is not valid in general: when N identical atoms are confined within a volume $V\ll\lambda^3$, where $\lambda$ is the wavelength of emitted radiation, the assumption of uncorrelated emitters is no longer valid and a closer reconsideration of the problem is necessary. It was shown that an ensemble of atoms coherently coupled to the quantum electromagnetic field, acts as a single quantum emitter, with a decay rate equal to $N A$, and an intensity of the emitted radiation proportional to $N^{2}$ \cite{persico,benedict}. This enhanced single-photon emission is known as \textit{superradiance}, and its physical origin is in the correlation (symmetric state) between the atomic dipoles, leading to a constructive interference in the emission of radiation.

The counterpart of superradiance is the so-called \textit{subradiance} \cite{dicke54,gross}, that occurs when the ensemble of atoms is prepared in a correlated antisymmetric state. In this case, a suppression of the emission intensity occurs, and the decay is totally inhibited. Contrarily to superradiance, subradiance arises from anticorrelations between the atomic dipoles, leading to a destructive interference in the emission of radiation. While superradiant states are affected by decoherence, subradiant states are free-decoherence robust states, and for these reasons they are considered promising for realization of high-performance quantum processors in quantum information technologies~\cite{petrosyan}.

Superradiance and subradiance have been investigated in a variety of systems, including atoms~\cite{pavolini,braggio}, trapped ions~\cite{devoe}, quantum dots~\cite{scheibner} coupled to various environments, such as cavities~\cite{temnov,pan}, waveguides~\cite{martin-cano,fleury}, and photonic crystals~\cite{ghhmypk15}.

Very recently, the influence of a perfect reflector on the cooperative spontaneous emission process of two atoms located nearby has been discussed~\cite{palacino}.
The effect of a surface or a structured environment, or of an external static electric field on other radiative processes, such as dispersion or resonance interactions between atoms, have been recently studied~\cite{incardone, notararigo,fiscelli, fiscelli1,shahmoon,shahmoon1,fiscelli2}.

Most of these studies concern with a static environment.
In this paper, we consider a different and more general situation, specifically we discuss the influence of a dynamical (i.e. time-dependent) environment on the cooperative emission of two correlated identical atoms located nearby.

Generally speaking, a dynamical environment can be realized by changing periodically the magneto-dielectric properties of the material or by a mechanical motion of macroscopic objects, such as a reflecting mirror or the cavity walls.
These systems, for example vibrating cavities or oscillating mirrors, have been extensively explored in connection with the dynamical Casimir and Casimir-Polder effect~\cite{dodonov, braggio1, dodonov1,antezza}. Also, dynamical cavities have been simulated in circuit QED~\cite{bosco}.

Recent investigations have shown that the presence of a dynamical environment can give additional possibilities (not present in the case of a static environment) to manipulate and control radiative properties of atoms or molecules coupled to a quantum field.
For example the spontaneous emission of an excited atom located near a perfectly reflecting plate that oscillates adiabatically has been recently discussed~\cite{glaetze, ferreri}, and it has been shown that the motion of the mirror significantly affects the atomic decay rate, as well as the spectrum of the emitted radiation, exhibiting the presence of two lateral and almost symmetrical peaks, not present in the case of a static boundary~\cite{ferreri}. Similar results
were also obtained in the case of an excited atom embedded in a dynamical photonic crystal, when its transition frequency is close to the photonic band edge of the photonic crystal~\cite{calajo}. Here, the presence of a time-modulated photonic band-gap gives rise to two lateral peaks in the spectrum of radiation emitted. These lateral peaks are asymmetric due to the rapidly varying local density of states at the edge of the gap.
Furthermore, the time-dependent resonance interaction between atoms, the dynamical Casimir-Polder interaction between atoms or between an atom and a mirror, have been investigated during the dynamical self-dressing process of the system, starting from a nonequilibrium configuration; it has been shown that forces usually attractive can become repulsive in non-equilibrium situations~\cite{rizzuto,vasile,messina1,armata,barcellona, barcellona1,haakh}. These results show the striking potentialities of time-dependent environments and nonequilibrium configurations for manipulating a variety of radiative processes.

In this paper, we consider two identical atoms, one in the ground state and the other in the excited state, prepared in a correlated {\em Bell}-type state (symmetric or antisymmetric), while the electromagnetic field is in its vacuum state. In the Dicke model, these states are the well-known superradiant and subradiant states, respectively~\cite{dicke54}.
We assume that the two atoms are located near a perfectly reflecting mirror that oscillates adiabatically along a prescribed trajectory, and
we investigate the effects of the mirror's motion on the cooperative spontaneous decay, the spectrum emitted by the two quantum emitters, and their decay rate.
We suppose that the reflecting plate oscillates adiabatically along a sinusoidal trajectory. Under these assumptions, the field mode functions, satisfying the boundary conditions at the mirror surface at any time, are time-dependent.
Using time-dependent perturbation theory, we investigate the spectrum of the emitted radiation, and the cooperative decay rate of the two-atom system.
We show that the adiabatic motion of the mirror modifies the physical features of the spectrum of the radiation emitted. In particular, we find the presence of two symmetric side peaks in the spectrum, not present in the case of a static mirror, and separated by the central peak by the mirror's oscillation frequency.
We also evaluate the transition rate to the collective ground state of the two-atom system, in both cases of the superradiant (symmetric) and subradiant (antisymmetric) state, and show that it depends on the interatomic separation and the time-dependent atom-plate distances.
We also find that the motion of the mirror can cause a significant enhancement or suppression of superradiance of the two quantum emitters, depending from the specific configuration of the system, with respect to the cases of a mirror at rest or atoms in the unbounded space. These results show how a dynamical environment can influence the physical features of the superradiant and subradiant emission by the two correlated atoms, that can be enhanced or inhibited compared to the case of atoms in the vacuum space or near a static boundary.
In general, this further confirms that a dynamical (i.e. time-modulated) environment can give new possibilities to control, manipulate and also activate or inhibit radiative processes of atoms and molecules nearby, such as the cooperative spontaneous emission by two correlated atoms. It suggests that also other radiative processes, such as the resonance interaction and the energy transfer between atoms or molecules, can be tailored exploiting a dynamical environment.

The paper is organized as follows. In Section \ref{sec:2}, we introduce our system, and investigate the spectrum of the radiation emitted by the two-atom system, and discuss its main physical features (some technical points on our model are in the Appendix). In Section \ref{sec:3} we investigate the collective decay rate of the two quantum emitters in the presence of the oscillating mirror. Section \ref{sec:4} is devoted to our concluding remarks.

\section{\label{sec:2} Spectrum of the radiation emitted by two entangled atoms near an oscillating mirror}

Let us consider two atoms, labeled as A and B, located in the half-space $z>0$ near an infinite perfectly conducting plate, modeled as two-level systems with atomic transition frequency $\omega_0$, and interacting with the electromagnetic field in the vacuum state.
We suppose that the mirror oscillates with an angular frequency $\omega_p$, along the $z$ direction with the trajectory
$a(t) = a \sin(\omega_p t)$, where $a$ is the oscillation amplitude of the plate around its average position $z=0$.

Let us suppose that the two identical two-level atoms are initially prepared in a symmetric or antisymmetric entangled state, i.e.
\begin{eqnarray}
\label{eq:2}
\vert\phi\rangle_{\pm}=\frac{1}{\sqrt{2}}(\vert e_A,g_B\rangle\pm\vert g_A, e_B\rangle)\, ,
\end{eqnarray}
and that the quantum field is in its vacuum state. Thus, the initial state of the system at time $t=0$ is
\begin{eqnarray}
\label{eq:3}
\vert i \rangle_{\pm}=\vert\phi\rangle_{\pm}\vert vac\rangle\, .
\end{eqnarray}
The sign $\pm$ in \eqref{eq:2} refers to the symmetric or antisymmetric state respectively, $\vert vac\rangle$ is the vacuum state of the electromagnetic field, while $\vert e_{A(B)}\rangle$ ($\vert g_{A(B)}\rangle$) indicates the excited (ground) state of atom $A (B)$. In the states \eqref{eq:2} the excitation is delocalized between the two atoms. In the Dicke model, these states are the so-called superradiant and subradiant states, respectively.
They can be realized experimentally with actual techniques~\cite{filipp,mlynek}.
Symmetric (antisymmetric) states are also at the origin of the resonant interaction energy, which is a second-order interaction between correlated atoms~\cite{craig}.

\begin{figure}[!htbp]
\centering\includegraphics[width=8.2 cm]{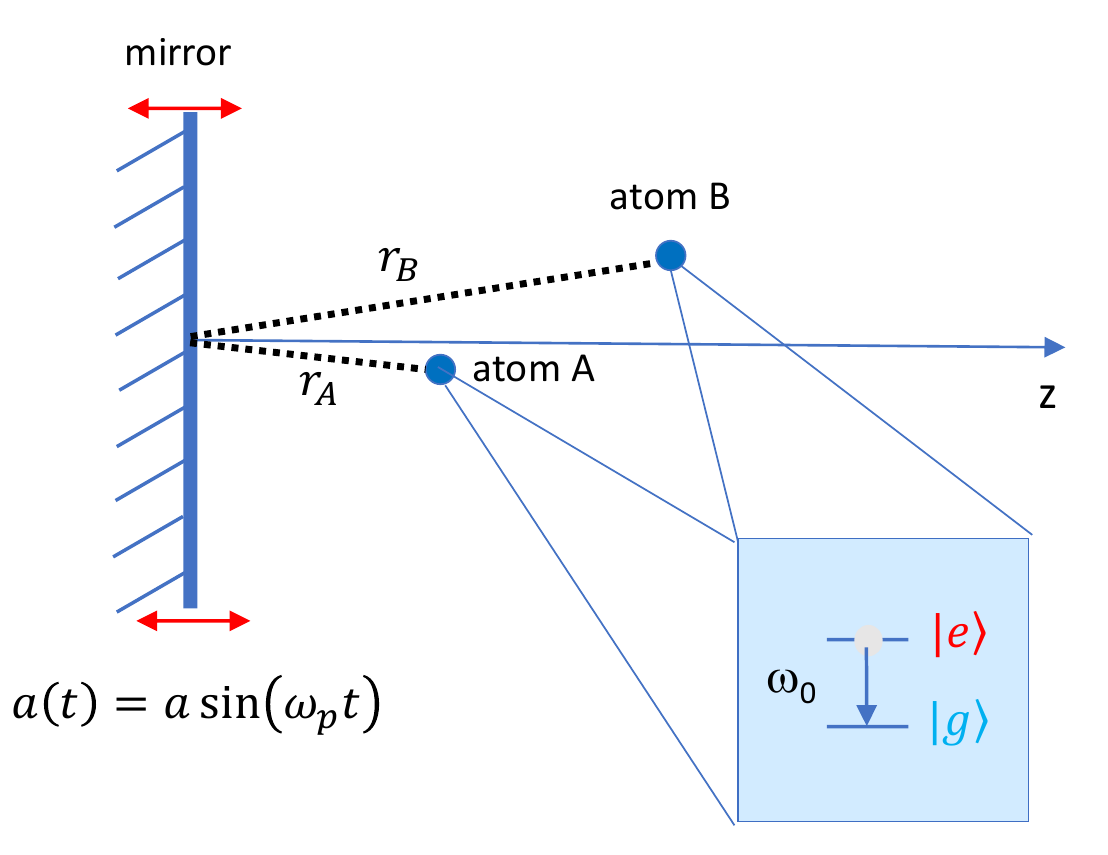}
\caption{Sketch of the system: two atoms, modeled as two-level systems with transition frequency $\wo$, are placed in front of an oscillating mirror. The atomic dipole moment of each atom can be oriented parallel or perpendicular to the oscillating reflecting plate.}
\label{fig:1}
\end{figure}

Our physical system is displayed in Figure~\ref{fig:1}.

We assume that the oscillation frequency $\omega_p$ of the plate is much smaller than the atomic transition frequency $\wo$ of both atoms, and of the inverse of the time taken by the photon emitted by one of the two atoms, to reach the other atom after reflection on the mirror ($\omega_p\ll c/r_A, c/r_B, c/(r_A+r_B)$, where $r_{A/B}$ is the average atom-plate distance of each atom from the mirror). Under these assumptions, we can neglect real photons emission by dynamical Casimir effect, and investigate the collective spontaneous emission by the two correlated atoms in the adiabatic approximation. These assumptions are fully verified by typical values of the relevant parameters of the system, for example $\omega_p \sim 10^9 \,$s$^{-1}$, $\omega_0 \sim 10^{15} \,$s$^{-1}$, and an atom-plate average distance of the order of $10^{-6}\,$m, achievable in the laboratory.
We stress that such a system is experimentally feasible, using a dynamical mirror, that is a slab of semiconductor material whose dielectric properties are modulated in time for simulating the oscillating mirror~\cite{braggio1, antezza}, and keeping the atoms at a fixed position exploting atomic trapping techniques~\cite{wieman}.

We write the Hamiltonian of our system in the Coulomb gauge and in the multipolar coupling scheme, within the dipole approximation~\cite{compagno,craig,salam,passante18}:
\begin{equation}
\label{eq:4}
H=\hbar\wo (S^A_z+S^B_z)+ \skj\hbar\wk \akjd\akj+H_I ,
\end{equation}
where $S_z=\frac{1}{2}(\vert e\rangle \langle e \vert - \vert g\rangle \langle g \vert)$ is the pseudospin atomic operator, $\akj$ ($\akjd$) are the bosonic annihilation (creation) operators for photons with wave vector $\bm{k}$ and polarization $j$, and $H_I$ is the interaction Hamiltonian, given by

\begin{equation}
\label{eq:5}
H_I = -(S_+^A+S_-^A)\bmu_A^{eg}\cdot{\bf E}(\br_A)-(S_+^B+S_-^B)\bmu_B^{eg}\cdot{\bf E}(\br_B).
\end{equation}

Here,
$S_+=\vert e \rangle \langle g \vert$, $S_-=\vert g \rangle \langle e \vert$ are atomic pseudospin operators, and
$\bmu_{A(B)}^{eg}= \langle e_{A(B)} \vert \hat{\bmu}_{A(B)}\vert g_{A(B)}\rangle$ are the matrix elements, assumed real, of the atomic dipole moment operator $\hat{\bmu}_{A(B)}$ of atom $A$ ($B$) between its excited and ground state.
${\bf E}(\br_{A(B)})$ is the electric field operator at the atomic position $\br_{A(B)}$; it includes a time dependence as a consequence of the motion of the conducting wall, as discussed in detail in the Appendix.

The orientation of the atomic dipole moment is determined by the specific atomic states $\lvert e \rangle$ and $\lvert g \rangle$ taken in our two-level model. For example, if the excited state is one of the three degenerate states with $n=2, \, l=1, \, m=0,\pm 1$ of the hydrogen atom, and the ground state is the state with $n=1, \, l=0, \, m=0$ ($n$ is the principal quantum number, $l$ the orbital quantum number, and $m$ the magnetic quantum number), then the excited state with $m=0$ gives a dipole matrix element along $z$ (perpendicular to the wall), while specific linear combinations of the states with $m=1$ and $m=-1$ give a dipole along $x$ or $y$, that is parallel to the wall.

In general, the presence of time-dependent boundary conditions leads to introducing new creation and annihilation field operators, related to the old ones by a Bogoliubov transformation~\cite{dodonov}, and to time-dependent mode functions, satisfying the appropriate time-dependent boundary conditions.
However, in the present case of an adiabatic motion of the mirror as defined above, the field operators instantaneously follow the mirror's motion, and the creation and annihilation operators are the same of the static-wall case; also, we can set the usual boundary conditions for the electromagnetic field in the reference where the wall is instantaneously at rest, and then go back to the laboratory frame by the appropriate time-dependent space translation. Thus, the field annihilation and creation operators remain the same as in the static case, and the mirror's motion is entirely included in the field modes $(\omega_pa/c \ll 1)$.
The expression for the field operator appearing in \eqref{eq:5} with the adiabatically moving mirror, as well as the instantaneous field modes and relevant expressions for the sum over polarizations, are discussed in the Appendix.

Thus, the time dependence of the interaction Hamiltonian will be made explicit in the mode functions only (see the Appendix), while, as mentioned, the field annihilation and creation operators are the same as in the static case; in other words, in our adiabatic approximation, the atoms locally interact with the vacuum field fluctuations that instantaneously follow the motion of the mirror. In general, in dealing with our system, we can adopt two different points of view: with respect to the laboratory frame, where both atoms are at rest and the plate oscillates along a prescribed trajectory, or in the reference frame comoving with the mirror.
In this paper, we will adopt the laboratory frame.
Taking into account our hypothesis of an adiabatic motion of the mirror, the field vacuum state in our initial state \eqref{eq:3}, at $t=0$, is that relative to the instantaneous position of the mirror at that time, and it is independent of its previous motion (in fact, nonadiabatic effects such as photon emission by dynamical Casimir effect, or atomic excitation by dynamical Casimir-Polder effect \cite{antezza}, are negligible in our hypothesis). Mathematically, it is defined by $\akj \vert vac \rangle =0$, where the annihilation operator $\akj$ does not depend from time, because all time dependence relative to the wall's motion is embedded in the field modes.

Using the interaction Hamiltonian \eqref{eq:5}, a straightforward application of first-order time-dependent perturbations theory yields the transition amplitude from the initial entangled state \eqref{eq:3} to the state $\vert g_A,e_B,1_{\bk ,j} \rangle$ (both atoms in their ground state, and one photon emitted in the mode $(\bk ,j))$
\begin{eqnarray}
\label{trans-ampl}
c(\bk j,t) &=& \sqrt{\frac {\pi ck}{\hbar V}} \int_0^t dt' e^{i(\wk -\wo )t'}  \nonumber \\
&\ & \times  \left[\bmu_A^{eg} \cdot {\bf E}(\br_A,t') \pm \bmu_B^{eg} \cdot{\bf E}(\br_B,t') \right] ,
\end{eqnarray}
where the $\pm$ sign refers to the superradiant or subradiant state of Eq.~\eqref{eq:2}.

The probability that the system, initially prepared in the correlated state \eqref{eq:3}, decays at time $t$ to the collective atomic ground-state, emitting a photon with wavevector ${\bk}$ and polarization $j$, is then given by

\begin{widetext}
\begin{eqnarray}
\label{eq:7}
&\ &\lvert c(\bk j,t)\rvert^2 = \frac{\pi c k}{\hbar V}\int_0^{t}\int_0^{t}d\tp d\tpp \Big\{\bmu_A^{eg}\cdot\bfm_{\bk j}(\br_A,\tp)\bmu_A^{eg}\cdot\bfm_{\bk j}(\br_A,\tpp)+\bmu_B^{eg}\cdot\bfm_{\bk j}(\br_B,\tp )\bmu_B^{eg}\cdot\bfm_{\bk j}(\br_B,\tpp )\nonumber\\
&\ & \ \ \pm \Big[\bmu_A^{eg}\cdot\bfm_{\bk j}(\br_A,\tp )\bmu_B^{eg}\cdot\bfm_{\bk j}(\br_B,\tpp )+\bmu_A^{eg}\cdot\bfm_{\bk j}(\br_A,\tpp )\bmu_B^{eg}\cdot\bfm_{\bk j}(\br_B,\tp )\Big]\Big\} e^{i(\wk-\wo)(\tpp-\tp)} .
\end{eqnarray}
\end{widetext}

The first two terms in the right-hand side of Eq.~\eqref{eq:7} are related to the probability that each atom independently decays by emitting a photon; on the contrary, the contribution inside the square bracket is an {\em interference} term, and it is responsible of the superradiant or subradiant behavior of the two-atom system.

From Eq.~\eqref{eq:7} we can obtain the frequency spectrum of the radiation emitted by the two
atoms, that is the emission probability for unit frequency, by taking the sum over polarization and the integration over the directions of $\bk$ as

\begin{equation}
\label{eq:8}
P(\wk,t)=\frac{V}{(2\pi)^3} \frac {\wk^2}{c^3} \int \! d\Omega \sum_j \lvert c(\bk j,t)\rvert^2 ,
\end{equation}

where $V$ is the quantization volume, and $\Omega$ the solid angle.
The integration over the directions of $\bk$ that we will explicitly perform in the following is on the full $4\pi$ solid angle, because our field modes, given in the Appendix, allow positive and negative values of the components of $\bk$.

We perform the sum over polarizations $j=1,2$, using the relation \eqref{eq:9} given in the Appendix, that extends to our adiabatic dynamical case the expression obtained in \cite{power82} for the static case. For convenience, we report here this relation
\begin{eqnarray}
\label{eq:9a}
&\ & \int \! d\Omega \sum_j [\bfm_{\bk j}(\br_u,\tp )]_\ell[\bfm_{\bk j}(\br_v,\tpp )]_m\nonumber\\
&\rightarrow& \int \! d\Omega \, \Re \Big\{(\delta_{\ell m}-{\hat \bk_\ell}{\hat \bk_m})e^{i\bk\cdot(\br_u-\br_v)}\nonumber\\
&\ & -\sigma_{\ell p}(\delta_{p m}-{\hat \bk_p}{\hat \bk_m})e^{i\bk\cdot(\br_u(\tp)-\sigma\br_v(\tpp))} \Big\}\, ,
\end{eqnarray}
where $\Re$ indicates the real part.

We stress that the relation \eqref{eq:9a} is valid only in our adiabatic approximation in the laboratory frame, that is when the electromagnetic field operators instantaneously follow the motion of the plate.
The first term in \eqref{eq:9a} is a free-space contribution,  and it is time-independent because the two atoms are fixed in space. On the contrary, the second term takes into account the presence of the oscillating mirror through the reflection matrix $\sigma = \text{diag}(1,1,-1)$ introduced in (\ref{eq:10}), and, in our adiabatic approximation, depends on the instantaneous time-dependent atom-mirror and atom-image distances (see the presence of the $\sigma$ reflection matrix).

The second term in Eq. \eqref{eq:9a} can be written as
\begin{equation}
\label{eq:11}
e^{i\bk\cdot(\br_u(\tp)-\sigma\br_v(\tpp))}=e^{i\bk\cdot\bar{\bR}_{uv}-{i\bk\cdot\bf a}[\sin(\omega_p \tp)+\sin(\omega_p \tpp)]} ,
\end{equation}
where $\bar{\bR}_{uv}=\br_u-\sigma\br_v$.
For a single atom, say $A$, $u=v=A$, and $\bar{\bR}_{A}=\br_A-\sigma\br_A$ represents the distance of atom $A$ from its image through the mirror; on the other hand, $\bar{\bR}_{AB}=\br_A-\sigma\br_B$, is the distance of one atom (say $A$) from the image of the other atom (say $B$).
For small oscillation amplitudes, such that $a\ll R_{A(B)},\bar{R}_{A(B)},\bar{R}_{AB}$, we can perform a series expansion of the exponential function in \eqref{eq:11} in powers of $a$, obtaining
\begin{eqnarray}
\label{eq:12}
&\ &e^{i\bk\cdot(\br_u(\tp )-\sigma\br_v(\tpp ))}\simeq e^{i\bk\cdot\bar{\bR}_{uv}}\biggl[1- i (\bk\cdot\hat{\bf n}) a \nonumber\\
&\ &\times (\sin(\omega_p \tp)+\sin(\omega_p \tpp)) - \frac 1 2 (\bk\cdot\hat{\bf n})^2 a^2\nonumber\\
&\ &\times (\sin(\omega_p \tp)+\sin(\omega_p \tpp))^2+....\bigg]\, ,
\end{eqnarray}
where $\hat{\bf n}=(0,0,1)$ is the unit vector orthogonal to the oscillating plate.
We can now substitute the relations \eqref{eq:12} and \eqref{eq:9a} into \eqref{eq:7}, and integrate over time. Taking into account only terms up to the second order in the oscillation amplitude $a$, after some algebra we get

\begin{eqnarray}
\label{eq:13}
\int \! d\Omega \sum_j \lvert c(\bk j,t)\rvert^2 &\simeq&
g_A(\omega_k,t) + g_B(\omega_k,t)
\nonumber\\
&\ & \pm g_{AB}(\omega_k,t)
\end{eqnarray}

where
\begin{widetext}
\begin{eqnarray}
\label{eq:14}
&\ &
g_{A(B)}(\omega_k,t) =
\frac{\pi c k}{\hbar V}(\bmu_{A(B)}^{eg})_{\ell}(\bmu_{A(B)}^{eg})_{m}
 \Re \int \! d\Omega
\Big[
(\delta_{\ell m}-{\hat \bk_\ell}{\hat \bk_m}) h_0(\omega_k-\omega_0,t) - \sigma_{\ell p}(\delta_{p m}-{\hat \bk_p}{\hat \bk_m})e^{i\bk\cdot{\bar\bR}_{A(B)}}  \nonumber\\
&\ & \times \Big( h_0(\wk-\wo,t) - i (\bk\cdot\hat{\bf n}) a h_1(\wk-\wo,\omega_p,t) - (\bk\cdot\hat{\bf n})^2 \frac{a^2}{2} (h_2(\wk-\wo,\omega_p,t)+h_3(\wk-\wo,\omega_p,t)) \Big) \Big]
\end{eqnarray}
\end{widetext}
are the single-atom contributions ($\Re$ indicates the real part), and
\begin{widetext}
\begin{eqnarray}
\label{eq:15}
&\ &
g_{AB}(\omega_k,t) = \frac{4\pi c k}{\hbar V}(\bmu_{A}^{eg})_{\ell}(\bmu_{B}^{eg})_{m} \Re \int \! d\Omega
\Big[(\delta_{\ell m}-{\hat \bk_\ell}{\hat \bk_m})e^{i\bk\cdot{\bR}_{AB}} h_0(\omega_k-\omega_0,t) - \sigma_{\ell p}(\delta_{p m}-{\hat \bk_p}{\hat \bk_m})e^{i\bk\cdot{\bar\bR}_{AB}}
\nonumber\\
&\ &\times \Big( h_0(\wk-\wo,t)- i (\bk\cdot\hat{\bf n}) a\, h_1(\wk-\wo,\omega_p,t) - (\bk\cdot\hat{\bf n})^2 \frac{a^2}{2} (h_2(\wk-\wo,\omega_p,t)+h_3(\wk-\wo,\omega_p,t))\Big) \big] .
\end{eqnarray}
\end{widetext}
is the {\em interference} term, yielding the cooperative effects. In the expressions \eqref{eq:14} and \eqref{eq:15}, we have introduced the following functions

\begin{eqnarray}
&\ & h_0(\wk-\wo,t)=\frac{\sin^2((\wk-\wo)t/2)}{((\wk-\wo)/2)^2} ,
\label{eq:16}
\end{eqnarray}

\begin{widetext}
\begin{eqnarray}
\label{eq:17}
&\ &h_1(\wk -\wo,\omega_p,t)=\sin(\omega_p t/2)\frac{\sin[(\wk -\wo )t/2]}{(\wk -\wo )/2}
\left(\frac{\sin[(\wk -\wo +\omega_p)t/2]}{(\wk -\wo +\omega_p)/2}
+\frac{\sin[(\wk -\wo -\omega_p)t/2]}{(\wk -\wo -\omega_p)/2}\right) ,
\end{eqnarray}

\begin{eqnarray}
\label{eq:18}
h_2(\wk -\wo,\omega_p,t)&=& \frac{\sin^2[(\wk -\wo +\omega_p)t/2])}{(\wk -\wo +\omega_p)^2/2}+\frac{\sin^2[(\wk -\wo-\omega_p)t/2]}{(\wk -\wo -\omega_p)^2/2} -\cos (\omega_p t) \nonumber \\
&\ & \times \frac {\sin[(\wk -\wo +\omega_p)t/2] \sin [(\wk -\wo -\omega_p)t/2)]}{(\wk -\wo +\omega_p)(\wk -\wo -\omega_p)/4} ,
\end{eqnarray}

\begin{eqnarray}
\label{eq:19}
h_3(\wk -\wo,\omega_p,t) &=& \frac{\sin^2[(\wk -\wo ) t/2]}{[(\wk -\wo )/2]^2}
-2\cos (\omega_p t) \frac {\sin [(\wk -\wo) t/2]}{\wk -\wo}\left( \frac {\sin [(\wk -\wo +2\omega_p)t/2]}{\wk -\wo +2\omega_p}\right.\nonumber\\
&\ &\left. + \frac{\sin [(\wk -\wo -2\omega_p)t/2]}{\wk -\wo -2\omega_p} \right) .
\end{eqnarray}
\end{widetext}

These functions give the behaviour of the emitted spectrum by the two-atom system, as a function of the mirror's oscillation frequency $\omega_p$ and the atomic transition frequency $\wo$. They are responsible of the qualitative features and changes (with respect to the fixed-mirror case) of the spectrum of the radiation emitted, due to the motion of the boundary. In fact, inspection of \eqref{eq:16}-\eqref{eq:19} clearly shows that, in addition to the usual central peak at $\wk=\wo$ (present also in the case of a static mirror), new lateral peaks at $\wk=\wo \pm \omega_p$ appear in the spectrum, due to the presence of energy denominators as $\wk -\wo \pm \omega_p$ in Eqs.~(\ref{eq:17}-\ref{eq:19}). These contributions are clearly related to the motion of the mirror, and vanish in the limit of a static boundary, namely when $a$ and/or $\omega_p$ vanish.

Substituting Eqs. (\ref{eq:13}-\ref{eq:15}) into (\ref{eq:8}), and separating the terms according to the order of the plate's oscillation amplitude $a$, some straightforward algebra finally yields the expression of the emission spectrum in the form

\begin{equation}
\label{eq:21}
P(\wk,t)=P^{(0)}(\wk,t)+P^{(1)}(\wk,t)+P^{(2)}(\wk,t) ,
\end{equation}
where $P^{(0)}(\wk,t)$ is the $0$-th order contribution, while $P^{(1)}(\wk,t)$ and
$P^{(2)}(\wk,t)$ give respectively the first- and second-order (in the mirror's oscillation amplitude $a$) modification to the spectrum consequent to the adiabatic motion of the mirror. Such contributions are
\begin{widetext}
\begin{eqnarray}
\label{eq:22}
P^{(0)}(\wk,t) &=&
\frac{ k^3 }{2\pi\hbar}
\sum_{u=A}^{B}(\bmu_u^{eg})_{\ell}(\bmu_u^{eg})_{m}\biggl[\frac 2 3 \delta_{\ell m}-\sigma_{\ell p}F_{m p}^{\bar{R}_u}\frac{\sin(k\bar{R}_u)}{k^3\bar{R}_u}\biggr]\frac{\sin^2((\wk-\wo)t/2)}{((\wk-\wo)/2)^2}\nonumber\\
&\ & \pm\frac{c k^3 }{\pi\hbar}(\bmu_A^{eg})_{\ell}(\bmu_B^{eg})_{m}\biggl[F_{\ell m }^{R_{AB}}\frac{\sin(kR_{AB})}{k^3 R_{AB}}-\sigma_{\ell p}F_{m p}^{\bar{R}_{AB}}\frac{\sin(k\bar{R}_{AB)})}{k^3\bar{R}_{AB}}\biggr]\frac{\sin^2((\wk-\wo)t/2)}{((\wk-\wo)/2)^2} ,
\end{eqnarray}

\begin{eqnarray}
\label{eq:23}
P^{(1)}(\wk,t)&=&
\frac{k^3 }{2\pi\hbar}
a\sigma_{\ell p}\biggl[\sum_{u=A}^B(\bmu_u^{eg})_{\ell}(\bmu_u^{eg})_{m} (\hat{\bf n}\cdot\nabla^{\bar{R}_u})F_{m p}^{\bar{R}_u}\frac{\sin(k\bar{R}_u)}{k^3\bar{R}_u}\pm 2(\bmu_A^{eg})_{\ell}(\bmu_B^{eg})_{m} (\hat{\bf n}\cdot\nabla^{\bar{R}_{AB}})F_{m p}^{\bar{R}_{AB}}\frac{\sin(k\bar{R}_{AB})}{k^3\bar{R}_{AB}}\biggr]\nonumber\\
&\ &\times h_1(\wk-\wo,\omega_p,t) ,
\end{eqnarray}

\begin{eqnarray}
\label{eq:24}
P^{(2)}(\wk,t) &=& -\frac{k^3 }{2\pi\hbar}
\frac{a^2}{2}\sigma_{\ell p} \biggl[\sum_{u=A}^B(\bmu_u^{eg})_{\ell}(\bmu_u^{eg})_{m}(\hat{\bf n}\cdot\nabla^{\bar{R}_u})^2 F_{m p}^{\bar{R}_u}\frac{\sin(k\bar{R}_u)}{k^3\bar{R}_u}\pm2(\bmu_A^{eg})_{\ell}(\bmu_B^{eg})_{m}(\hat{\bf n}\cdot\nabla^{\bar{R}_{AB}})^2 F_{m p}^{\bar{R}_{AB}}\frac{\sin(k\bar{R}_{AB})}{k^3\bar{R}_{AB}}\biggr]\nonumber\\
&\ &\times \Big[ h_2(\wk-\wo,\omega_p,t)+h_3(\wk-\wo,\omega_p,t) \Big] .
\end{eqnarray}
\end{widetext}
Here
\begin{eqnarray}
\label{eq:25}
F_{\ell m}^{r}=(-\delta_{\ell m}\nabla^2+\nabla_{\ell}\nabla_m)^r
\end{eqnarray}
is a differential operator acting on variable $r$, $R_{AB}=\lvert\br_A-\br_B\rvert$,
$\bar{R}_{A(B)}=\lvert \br_{A(B)}-
\sigma\br_{A(B)}\rvert$, $\bar{R}_{AB}=\lvert \br_A-\sigma\br_B\rvert$, and $\br_A$, $\br_B$ respectively being the positions of atoms A and B.

A comparison of these expressions with the analogous quantity for the static-mirror case, shows that the main difference is the presence of terms related to the oscillation frequency of the mirror, specifically two new lateral peaks in the spectrum at frequencies $\wk=\wo\pm\omega_p$. Their relative intensities are of the order of $a/\bar{R}_i$ (see Eq.~\eqref{eq:23}), and $(a/\bar{R}_i)^2$ (see Eq.~\eqref{eq:24}), and give a qualitative change of the spectrum.
We wish to point out that secondary lateral peaks at frequency $\wk=\wo\pm2\omega_p$, stemming from second-order terms in the expansion in $a$, are also present (as Eq. \eqref{eq:19} shows). They represent, at the order considered, a sort of nonlinear effect; however, within the range of validity of our approximations, they give a quite small contribution to the overall spectrum.

Our expression for $P(\wk,t)$ is valid for a generic geometric configuration of the two atoms with respect to the oscillating plate.
\begin{figure}[h]
\includegraphics[width=8.0 cm]{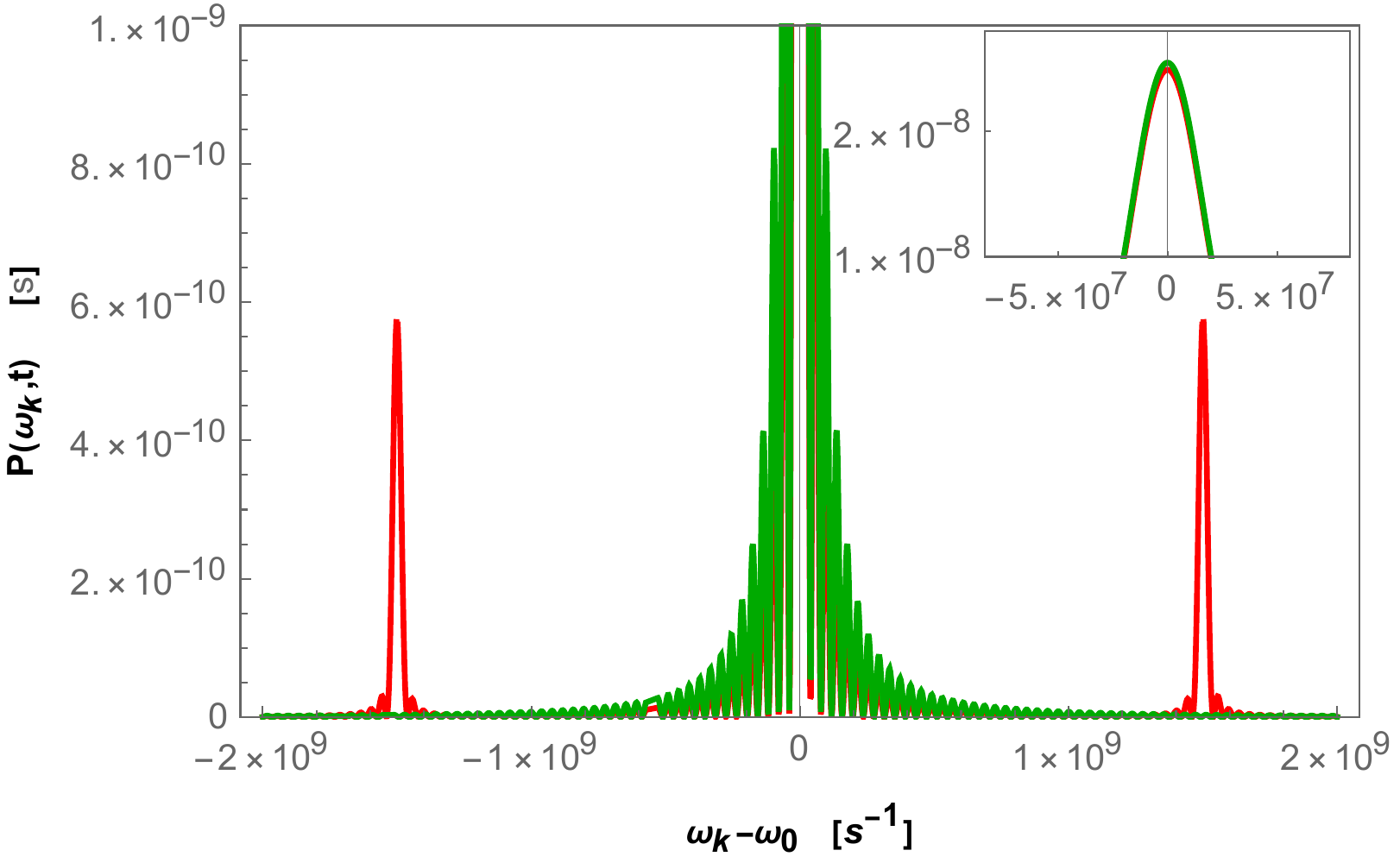}
\caption{Spectrum (scaled with respect to the total emission probability) emitted by the two-atom system, prepared in the correlated symmetric state, as a function of the detuning $\wk -\wo$, both in the static case (green line) and in the dynamical case, with the two lateral peaks (red line). The atoms are aligned perpendicularly to the mirror, with dipole moments along the $x$-axis (parallel to the plate). The figure clearly shows that the presence of a dynamical mirror  produces two lateral peaks (red line) shifted from the central peak by the mirror's modulation frequency.  The inset shows a zoom of the central peak in the two cases considered. Parameters are chosen such that $a=2\times 10^{-7}\,$m, $z^0_A = 10^{-6}\,$m, $z^0_B= 1.1\times10^{-6}\,$m, $\omega_p =1.5\times10^{9}\,$s$^{-1}$, $\omega_0 = 10^{15}\,$s$^{-1}$, $t =1.6\times10^{-7}\,$s, $\mu\sim10^{-30}$C$\cdot$m.}
\label{fig:2a}
\end{figure}
\begin{figure}[h]
\includegraphics[width=8.0 cm]{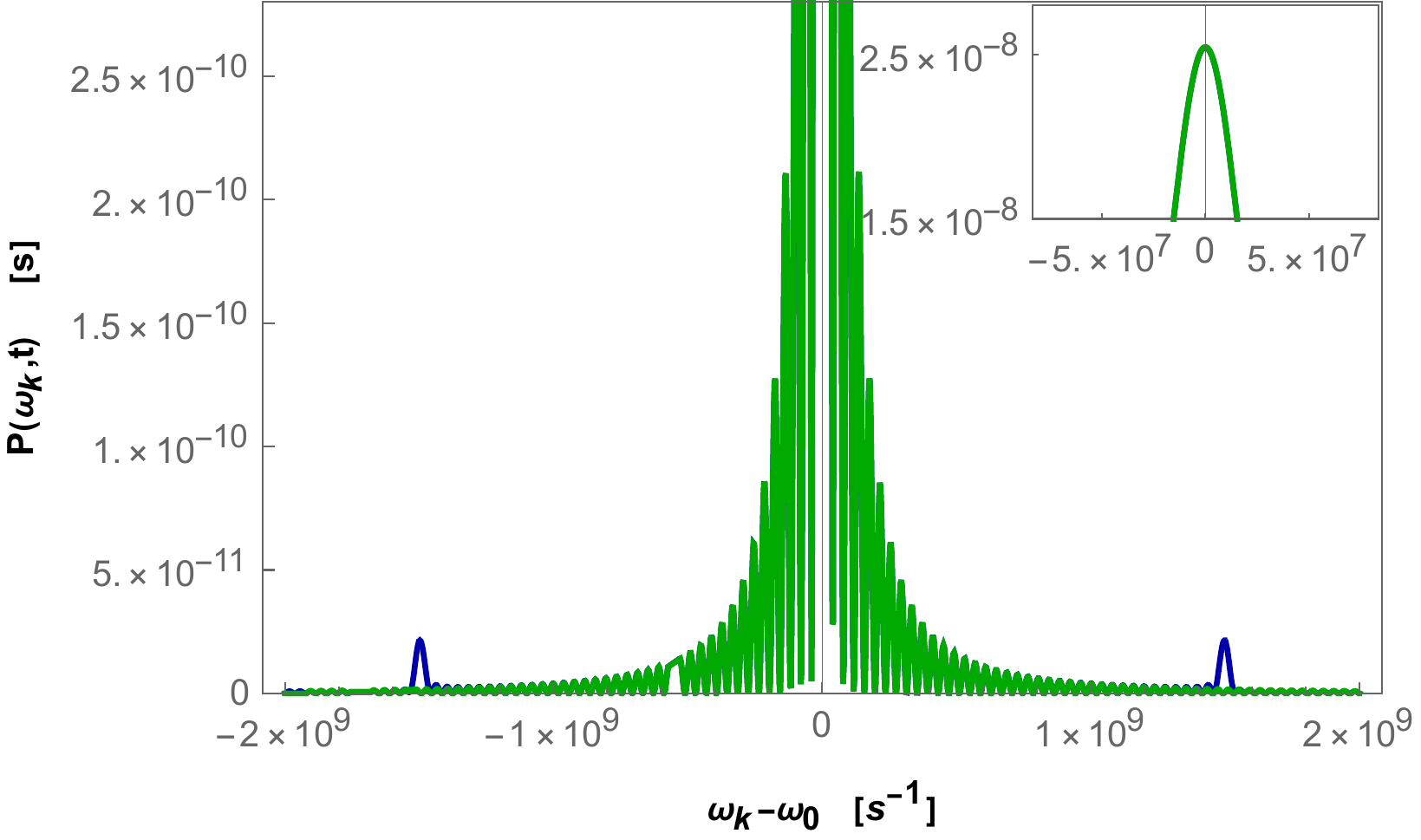}
\caption{Spectrum (scaled with respect to the total emission probability) emitted by the two-atom system, prepared in the correlated symmetric state, as a function of the detuning $\wk -\wo$, both in the static case (green line) and in the dynamical case with the two lateral peaks (blue line). The dipole moments are perpendicular to the plate (along the $z$-axis). As before, the presence of a dynamical mirror  produces two lateral peaks (blue line) shifted from the central peak by the mirror's modulation frequency.  The inset shows a zoom of the central peak in the two cases considered (the two curves practically overlap each other) . The numerical values of the parameters are the same as in the plot in Fig. \ref{fig:2a}.}
\label{fig:2b}
\end{figure}
\begin{figure}[h]
\includegraphics[width=8.0 cm]{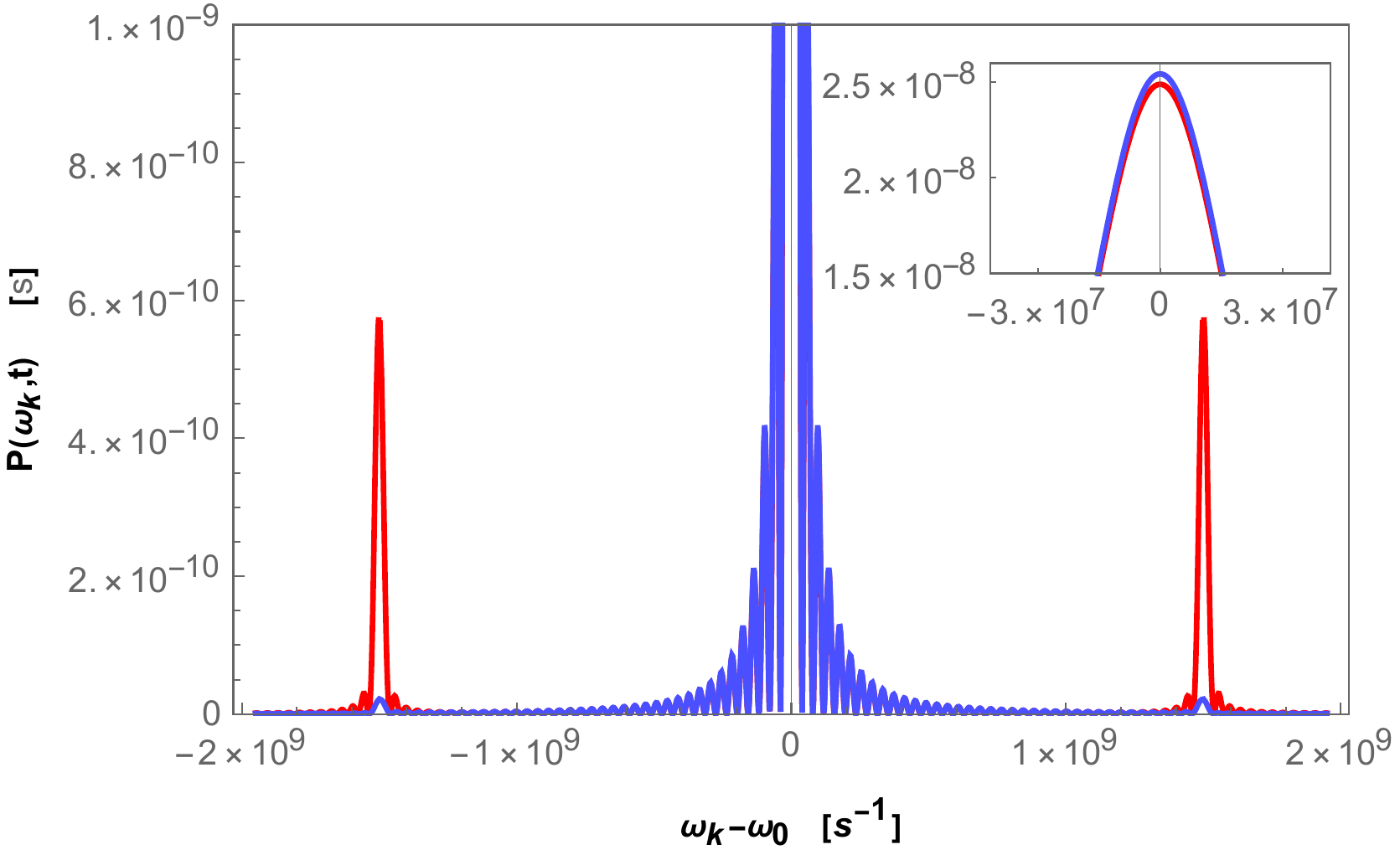}
\caption{Comparison between the emitted spectra (scaled with respect to the total emission probability) by the two-atom system, when the dipole moments are aligned parallel (red line) and perpendicular (blue line) to the plate. The figure shows that the lateral peaks in the emitted spectrum by dipole moments aligned along the $z$-axis are smaller than those obtained in the case of dipole moments oriented parallel to the mirror (along the $x$-axis). The inset shows a zoom of the central peak in the two cases considered. The numerical values of the parameters are the same as in Fig. \ref{fig:2a}.}
\label{fig:2c}
\end{figure}

In order to get a clear physical insight it is helpful to analyze $P(\wk,t)$ in the specific case of atoms aligned along the $z$ axis (i.e. perpendicularly to the mirror), for example when $\br_A=(0,0,z^0_{A})$ and $\br_B=(0,0,z^0_{B})$.
Figure \ref{fig:2a} shows the spectrum (scaled with respect to the total emission probability) in the symmetric case and in the limit of long times, as a function of the detuning $\wk -\wo$: the red line shows the dynamical case, while the green line shows the static-mirror case. As the figure shows,
the presence of the dynamical mirror determines the two symmetric lateral peaks shifted from the central peak by the modulation frequency. These two lateral peaks are symmetric with respect to the central peak, because the photonic density of states is essentially the same at the two frequencies. Analogous lateral peaks were found for a single two-level atom located near an oscillating mirror~\cite{ferreri}. A similar result is obtained for dipole moments aligned perpendicularly to the mirror, as figure \ref{fig:2b} shows. Interestingly, although the image dipole of $\bmu_{\perp}$ is
still $\bmu_{\perp}$, and a constructive interference between the atomic dipoles and their mirror images is expected, the intensity of the two lateral peaks in the emitted spectrum is smaller than that obtained in the case of dipole moments oriented parallel to the plate, as shown in figure \ref{fig:2c}. This effect seems to suggest that the oscillation of the mirror can induce a sort of destructive interference between the atomic dipoles and their images, oriented along the $z$-direction, parallel to the motion of the plate.

\begin{figure}[!hbpt]
\includegraphics[width=8.0 cm]{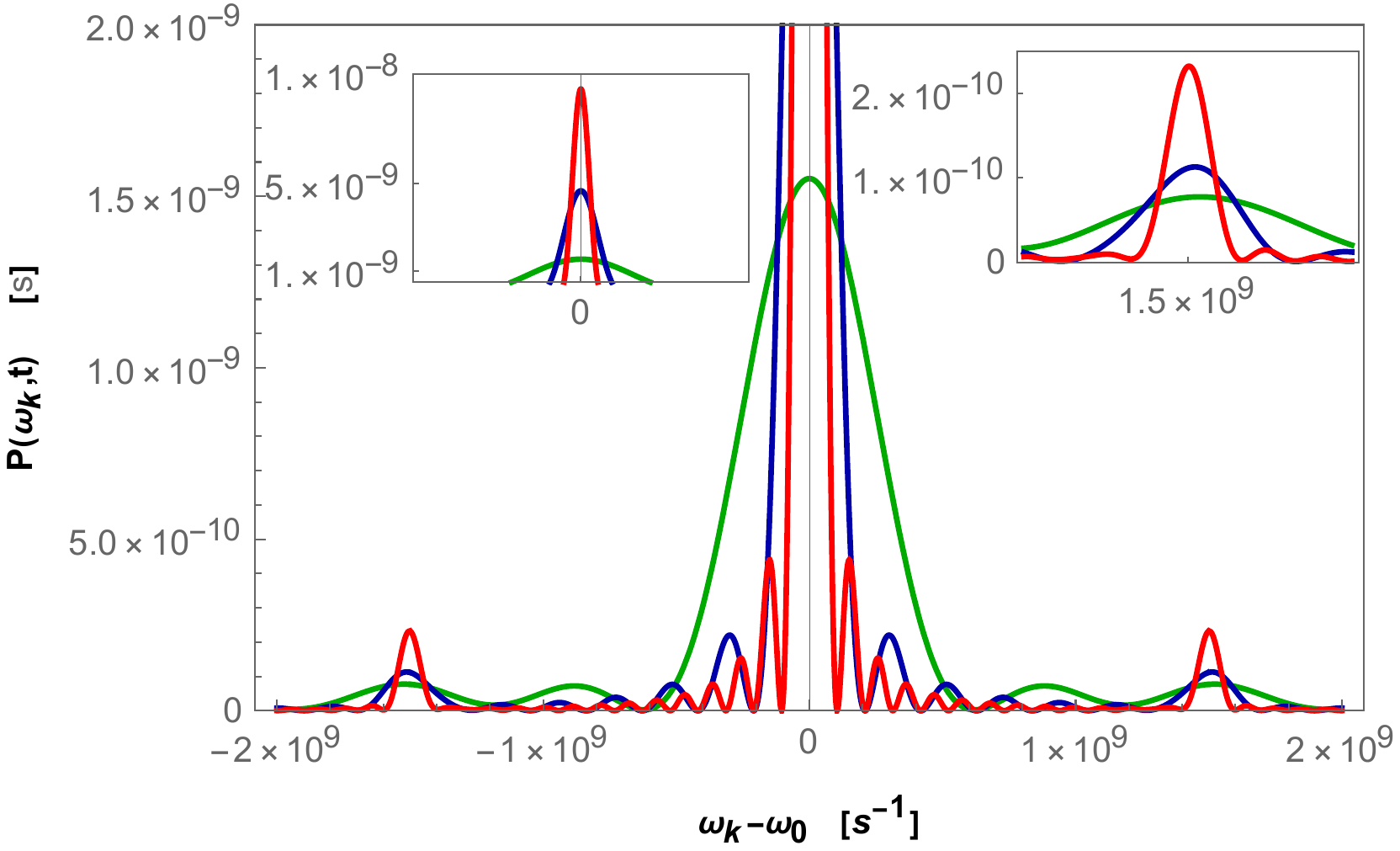}
\caption{Comparison between the emitted spectra (scaled with respect to the total emission probability) by the two-atom system, at different times. The dipole moments are aligned parallel to the plate. Continuous green line, blue line, and red line refer to the emitted spectrum at times $t=10^{-8}$s (continuous green), $t=3\times 10^{-8}$s (continuous blue), $t=6\times 10^{-8}$s (continuous red), respectively. The figure shows that the central peak (see inset on the left) and the two lateral peaks (see inset on the right) in the spectrum increase with time, as expected. The numerical values of the other parameters are the same as in Fig. \ref{fig:2a}.}
\label{fig:2d}
\end{figure}

We have also considered the emitted spectrum by the two-atom system at different times. The results obtained are illustrated in Fig. \ref{fig:2d}, for two atoms prepared in a symmetric superposition with dipole moments  oriented parallel to the oscillating mirror.  The figure shows that  the central and the lateral peaks increase with time, as expected.

Finally, we wish to stress that our results are in principle detectable using current experimental techniques; for example, for two hydrogen atoms and typical optical transitions, the natural linewidth is of the order of $ \sim 10^8 \,$s$^{-1}$; thus an oscillation frequency in the range $\omega_p \sim 10^9 - 10^{10} \,$s$^{-1}$, that can be currently obtained exploiting the technique of dynamical mirrors~\cite{braggio1,antezza}, is sufficient to resolve the lateral lines in the emitted spectrum.

\section{\label{sec:3}Collective spontaneous decay rate of the two-atom system}
We now evaluate the decay rate of the two-atom system to the ground state. This is obtained by integrating $P(\wk,t)$ over $k$, and then taking its time derivative,

\begin{eqnarray}
\label{eq:30}
&\ &\Gamma(t)= \frac{d}{d t}\int_{0}^{\infty} d\omega_k P(\wk,t) .
\end{eqnarray}
Since the functions $h_0(\omega_k-\omega_0,t)$ and $h_i(\omega_k-\omega_0,\omega_p,t)$ ($i=1,2,3$) are strongly peaked at $\omega_k\sim\omega_0$ and $\omega_k\sim\omega_0\pm\omega_p$, we can approximate the space-dependent functions in \eqref{eq:22}-\eqref{eq:24}, by their expressions in $k_0$ (taking also into account that $\omega_p\ll ck_0$) and then take them out of the integrals.
Taking into account only terms up to the first order in the expansion \eqref{eq:12} on the mirror's oscillation amplitude, a straightforward calculation gives
\begin{eqnarray}
\label{eq:31}
\Gamma(t)=\Gamma_A(t)+\Gamma_B(t)\pm\Gamma_{AB}(t) ,
\end{eqnarray}
where
\begin{widetext}
\begin{eqnarray}
\label{eq:32}
\Gamma_{A(B)}(t)=\frac{k_0^3}{\hbar}(\bmu_{A(B)}^{eg})_{\ell}(\bmu_{A(B)}^{eg})_{m}\biggl[\frac 2 3 \delta_{\ell m}-\sigma_{\ell p}F_{m p}^{\bar{R}_{A(B)}}\frac{\sin(k_0\bar{R}_{A(B)})}{k_0^3\bar{R}_{A(B)}}+2a\sin(\omega_p t)\sigma_{\ell p}(\hat{\bf n}\cdot\nabla^{\bar{R}_{A(B)}}) F_{m p}^{\bar{R}_{A(B)}}\frac{\sin(k_0\bar{R}_{A(B)})}{k_0^3\bar{R}_{A(B)}}\biggr] ,
\end{eqnarray}
\begin{eqnarray}
\label{eq:33}
\Gamma_{AB}(t)=\frac{2k_0^3}{\hbar}(\bmu_{A}^{eg})_{\ell}(\bmu_{B}^{eg})_{m}\biggl[F_{\ell m }^{R_{AB}}\frac{\sin(k_0R_{AB})}{k_0^3R_{AB}}-\sigma_{\ell p}F_{m p}^{\bar{R}_{AB}}\frac{\sin(k_0\bar{R}_{AB})}{k_0^3\bar{R}_{AB}}+2a\sin(\omega_p t)\sigma_{\ell p}(\hat{\bf n}\cdot\nabla^{\bar{R}_{AB}}) F_{m p}^{\bar{R}_{AB}}\frac{\sin(k_0\bar{R}_{AB})}{k_0^3\bar{R}_{AB}}\biggr] \, .
\end{eqnarray}
\end{widetext}

The expressions \eqref{eq:31}-\eqref{eq:33} are general, valid for a generic configuration of the two atoms with respect to the plate, and show oscillations of the decay rate with time, directly related to the adiabatic motion of the mirror. In fact, the emission rate of our system shows a term that oscillates in time by following the mirror's law of motion, of course. This is strictly related to our hypothesis of adiabatic motion of the boundary.
In order to discuss in more detail this result, similarly to what we did in the previous section, we analyze the specific case of atoms aligned along the $z$-direction, i.e. perpendicular to the reflecting plate. In this case of a perpendicular orientation we obtain
\begin{widetext}
\begin{eqnarray}
\label{eq:34}
&\ &\Gamma_{A}(t) = \frac{k_0^3}{\hbar}(\bmu_{A}^{eg})_{\ell}(\bmu_{A}^{eg})_{m}\biggl\{\frac 2 3 \delta_{\ell m}-\sigma_{\ell p}\biggl[-\left(\delta_{p m}-3({\hat{\bar{R}}_{A}})_{p}({\hat{\bar{R}}_{A}})_{m}\right)\left(\frac{\sin{k_0\bar{R}_{A}}}{k_0^3\bar{R}_{A}^3}-\frac{\cos{k_0\bar{R}_{A}}}{k_0^2\bar{R}_{A}^2}\right)\nonumber\\
&\ &+\left(\delta_{p m}-({\hat{\bar{R}}_{A}})_{p}({\hat{\bar{R}}_{A}})_{m}\right)\frac{\sin{k_0\bar{R}_{A}}}{k_0\bar{R}_{A}}\biggr]+\frac{2a\sin(\omega_p t)}{\bar{R}_{A}}\sigma_{\ell p}\biggl[\left(\delta_{p m}-({\hat{\bar{R}}_{A}})_{p}({\hat{\bar{R}}_{A}})_{m}\right) \cos{k_0\bar{R}_{A}}\nonumber\\
&\ & -2\left(\delta_{p m}-3({\hat{\bar{R}}_{A}})_{p}({\hat{\bar{R}}_{A}})_{m}\right)\frac{\sin k_0\bar{R}_{A}}{k_0\bar{R}_{A}}+3\left(\delta_{p m}-5({\hat{\bar{R}}_{A}})_{p}({\hat{\bar{R}}_{A}})_{m}\right)\left(\frac{\sin{k_0\bar{R}_{A}}}{k_0^3\bar{R}_{A}^3}-\frac{\cos{k_0\bar{R}_{A}}}{k_0^2\bar{R}_{A}^2}\right)\nonumber\\
&\ &-\left(\delta_{mz}{(\hat{\bar{R}}_{A}})_{p}+\delta_{pz}{(\hat{\bar{R}}_{A}})_{m}\right)\left(\frac{\sin{k_0\bar{R}_{A}}}{k_0\bar{R}_{A}}
+3\frac{\cos{k_0\bar{R}_{A}}}{k_0^2\bar{R}_{A}^2}-3\frac{\sin{k_0\bar{R}_{A}}}{k_0^3\bar{R}_{A}^3}\right)\biggr]\biggr\} ,
\end{eqnarray}
\end{widetext}
\begin{eqnarray}
\label{eq:35}
&\ &\Gamma_B(t)=(\Gamma_A(t) \,\,\, \text{with}\, A \rightarrow B) ,
\end{eqnarray}
\begin{widetext}
\begin{eqnarray}
\label{eq:36}
&\ &\Gamma_{AB}(t) = \frac{2k_0^3}{\hbar}(\bmu_{A}^{eg})_{\ell}(\bmu_{B}^{eg})_{m}\biggl\{\biggl[-\left(\delta_{\ell m}-3({\hat{R}_{AB}})_{\ell}({\hat{R}_{AB}})_{m}\right)\left(\frac{\sin{k_0 R_{AB}}}{k_0^3 R_{AB}^3}-\frac{\cos{k_0 R_{AB}}}{k_0^2 R_{AB}^2}\right)\nonumber\\
&\ &+\left(\delta_{\ell m}-({\hat{R}_{AB}})_{\ell}({\hat{R}_{AB}})_{m}\right)\frac{\sin{k_0 R_{AB}}}{k_0 R_{AB}}\biggr]+\sigma_{\ell p}\biggl[\left(\delta_{p m}-3({\hat{\bar{R}}_{AB}})_{p}({\hat{\bar{R}}_{AB}})_{m}\right)\left(\frac{\sin{k_0\bar{R}_{AB}}}{k_0^3\bar{R}_{AB}^3}-\frac{\cos{k_0\bar{R}_{AB}}}{k_0^2\bar{R}_{AB}^2}\right)\nonumber\\
&\ &-\left(\delta_{p m}-({\hat{\bar{R}}_{A}})_{p}({\hat{\bar{R}}_{AB}})_{m}\right)\frac{\sin{k_0\bar{R}_{AB}}}{k_0\bar{R}_{AB}}\biggr]+\frac{2a\sin(\omega_p t)}{\bar{R}_{AB}}\sigma_{\ell p}\biggl[\left(\delta_{p m}-({\hat{\bar{R}}_{AB}})_{p}({\hat{\bar{R}}_{AB}})_{m}\right) \cos{k_0\bar{R}_{AB}}\nonumber\\
&\ & -2\left(\delta_{p m}-3({\hat{\bar{R}}_{AB}})_{p}({\hat{\bar{R}}_{AB}})_{m}\right)\frac{\sin k_0\bar{R}_{AB}}{k_0\bar{R}_{AB}}+3\left(\delta_{p m}-5({\hat{\bar{R}}_{AB}})_{p}({\hat{\bar{R}}_{AB}})_{m}\right)\left(\frac{\sin{k_0\bar{R}_{AB}}}{k_0^3\bar{R}_{AB}^3}-\frac{\cos{k_0\bar{R}_{AB}}}{k_0^2\bar{R}_{AB}^2}\right)\nonumber\\
&\ &-\left(\delta_{mz}{(\hat{\bar{R}}_{AB}})_{p}+\delta_{pz}{(\hat{\bar{R}}_{AB}})_{m}\right)\left(\frac{\sin{k_0\bar{R}_{AB}}}{k_0\bar{R}_{AB}}
+3\frac{\cos{k_0\bar{R}_{AB}}}{k_0^2\bar{R}_{AB}^2}-3\frac{\sin{k_0\bar{R}_{AB}}}{k_0^3\bar{R}_{AB}^3}\right)\biggr]\biggr\} .
\end{eqnarray}
\end{widetext}

Expressions (\ref{eq:34}-\ref{eq:36}) show that the motion of the mirror yields new time-dependent terms of the order of $a/\bar{R}_{A/B}$ and $a/\bar{R}_{AB}$. We have neglected second-order terms in the perturbative expansion; this approximation is valid for small oscillation amplitudes with respect to other relevant length scales in the system, that is for $a\ll \bar{R}_{A/B},\bar{R}_{AB}$ and $a\ll k_0^{-1}$. For example, for $k_0 \sim 10^7 \,$m$^{-1}$, $R_{A/B} \sim 10^{-6} \,$m, and $a=10^{-8} \,$m, we have $a/\bar{R}_{A/B}, \, a/\bar{R}_{AB} \sim 10^{-1}$, \, $k_0 a \sim 10^{-1}$, and we neglect the second-order term proportional to $a^2$. The conditions above are within reach of currently achievable experimental techniques.
\begin{figure}
\includegraphics[width=8.0 cm]{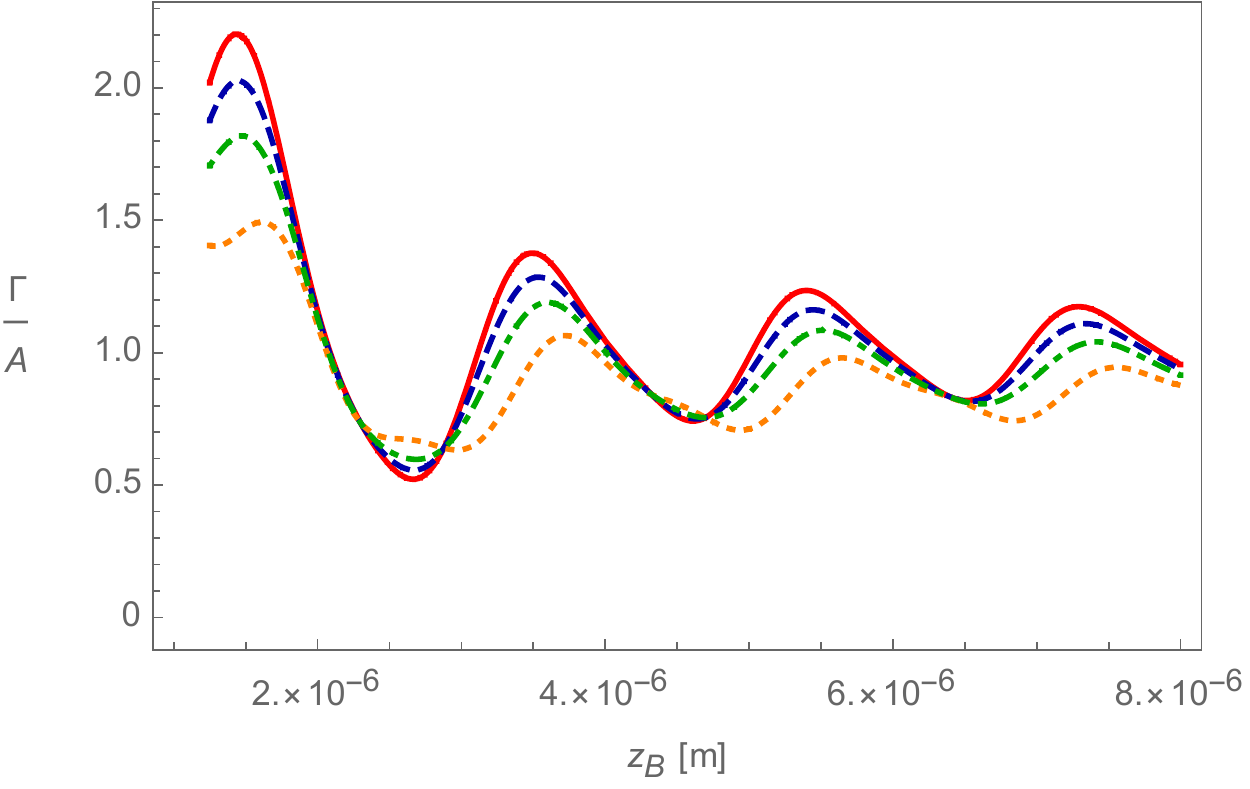}
\caption{Plot of the collective decay rate for two atoms prepared in the correlated symmetric state, at different times, as a function of the distance of atom $B$ from the mirror, when the atom $A$ is kept fixed ($R_A=z_A=1.25\times 10^{-6}$m). The atoms are aligned along the $z$ direction, with dipole moments parallel to the mirror. Continuous red line, blue dashed line, orange dotted line refer to two atoms near the oscillating mirror, at times $t=2\times 10^{-7}$s (continuous red), $t=2.3\times 10^{-7}$s (blue dashed), $t=2.4\times 10^{-7}$s (orange dotted), respectively. Dot-dashed green line refers to two atoms in the presence of a static mirror.
We have also used $a=2\times 10^{-7}\,$m, $\omega_p =1.5\times10^{9}\,$s$^{-1}$, $\omega_0 = 10^{15}\,$s$^{-1}$.}
\label{fig3}
\end{figure}
\begin{figure}
\includegraphics[width=8.0 cm]{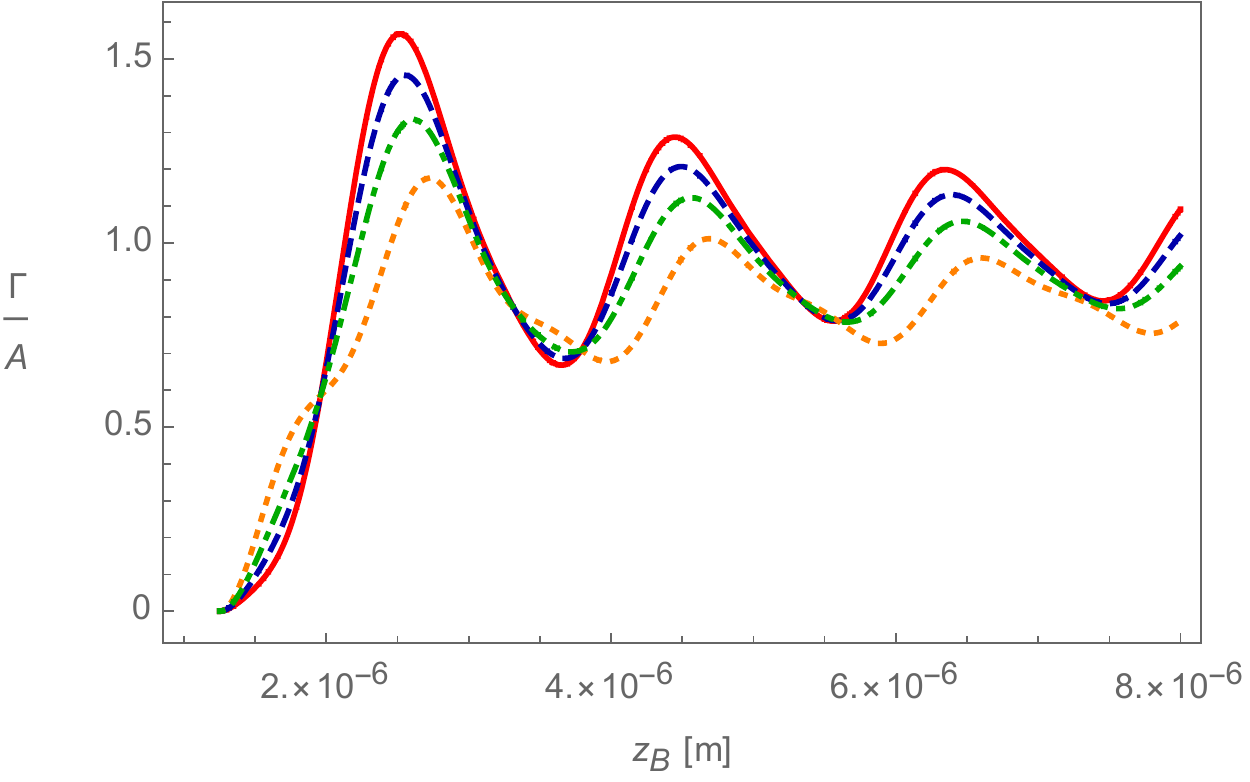}
\caption{Plot of the collective decay rate for two atoms prepared in the correlated antisymmetric state, at different times, as a function of the distance of atom $B$ from the mirror, when the atom $A$ is kept fixed ($R_A=z_A=1.25\times 10^{-6}$m). The atoms are aligned along the $z$ direction, with dipole moments parallel to the mirror, Continuous red line, blue dashed line, orange dotted line refer to two atoms near the oscillating mirror, at times $t=2\times 10^{-7}$s (continuous red), $t=2.3\times 10^{-7}$s (blue dashed), $t=2.4\times 10^{-7}$s (orange dotted), respectively. The green dot-dashed line refers to two atoms near a static mirror.
The other relevant parameters are the same of Fig. \ref{fig3}.}
\label{fig4}
\end{figure}

Figure \ref{fig3} shows the scaled (with respect to Einstein coefficient $A$, given in (\ref{eq:1})) collective decay rate at different times, as a function of the distance of atom $B$ from the mirror's average position, when atom $A$ is at a fixed position. The two atoms are aligned orthogonal to the mirror and prepared in the symmetric state. The dipole moments are parallel to the mirror. The figure shows that the decay rate oscillates in time, and that, at a given time, in specific distance ranges it can be increased (in the figure, see the red continuous line, blue dashed line and orange dotted line) with respect to the static-mirror case (green dot-dashed line); in other distance ranges, the opposite occurs. Analogous results are obtained in the case of atoms prepared in an antisymmetric configuration, as shown in figure \ref{fig4}.

In conclusion, our results show that the spectrum of the emitted radiation can be qualitatively changed exploiting the oscillation of the plate, and that the collective spontaneous emission can be controlled (enhanced or suppressed) by modulating in time the position of the mirror. This suggest the possibility to control also other radiative processes by modulated (time-dependent) environments, for example the resonance energy transfer between atoms or molecules.

\section{\label{sec:4}Conclusion}
In this paper, we have discussed the collective spontaneous decay of a system of two identical two-level atoms prepared in a correlated (symmetric or antisymmetric) {\em Bell-type} state, and located near an oscillating perfectly reflecting plate, in the adiabatic regime.
We have first discussed in detail the effect of the motion of the mirror on the spectrum of the radiation emitted by the two atoms, and then their collective spontaneous decay rate. We have shown that the motion of the mirror strongly affects the features of the spectrum, which exhibits, in addition to the usual peak at $\omega=\wo$, two new lateral peaks separated from the atomic transition frequency by the oscillation frequency of the plate, similarly to previous results for the single-atom decay~\cite{ferreri}.
We have also found that the decay rate to the collective ground-state is modulated in time, and can be increased or decreased, compared with the static-boundary case, according to time and atoms-wall distances, by exploiting the oscillating boundary. Our results show that modulated environments can give additional possibilities, with respect to fixed boundaries, to manipulate and tailor atomic radiative processes such as the cooperative spontaneous emission; also, they strongly indicate a similar possibility for other relevant radiative processes such as the energy transfer between two atoms, or the resonance interaction energy between correlated atoms. We will consider these physical systems in a future publication.

\appendix*
\section{Field operators and modes with the adiabatically moving mirror}
\label{App:1}

For a static mirror located at $z=0$, the electric field operator is given by
\begin{equation}
{\bf E}(\br )=i\sum_{\bk j} \sqrt{\frac{2\pi\hbar ck}{V}} \bfm_{\bk j}(\br )(\akjd-\akj) ,
\label{eq:6}
\end{equation}
where $\akj$ and $\akjd$ are annihilation and creation operators obeying usual boson commutation relations, and $\bfm_{\bk j}(\br )$ are the mode functions satisfying the appropriate boundary condition at the mirror's surface.
The annihilation and creation operators in \eqref{eq:6} are time-independent, because we are working in the Schr\"{o}dinger representation.
We start considering a cubic cavity of side $L$ with walls at $x = \pm L/2$, $y = \pm L/2$, $z=0$, $z=L$; in this case, we have~\cite{power82,ferreri}
\begin{eqnarray}
\label{eq:6a}
\left[\bfm_{\bk j}(\br )\right]_x&=&\sqrt{8}(\ekj )_x\cos\left[k_x\left(x+\frac L 2\right)\right]\nonumber\\
&\ &\times \sin\left[k_y\left(y+\frac L 2\right)\right]\sin\left[k_z z\right] ,\\
\label{eq:6b}
\left[\bfm_{\bk j}(\br )\right]_y&=&\sqrt{8}(\ekj )_y\sin\left[k_x\left(x+\frac L 2\right)\right]\nonumber\\
&\ &\times \cos\left[k_y\left(y+\frac L 2\right)\right]\sin\left[k_z z \right] ,\\
\label{eq:6c}
\left[\bfm_{\bk j}(\br )\right]_z&=&\sqrt{8}(\ekj )_z\sin\left[k_x\left(x+\frac L 2\right)\right]\nonumber\\
&\ &\times\sin\left[k_y\left(y+\frac L 2\right)\right]\cos\left[k_z z\right] ,
\end{eqnarray}
where $\ekj \, (j=1,2)$ are unit polarization vectors, assumed real, satisfying $\ekj \cdot \hat{{\bf e}}_{\bk j'} = \delta_{jj'}$ and $\ekj \cdot \bk =0$. In the limit $L\rightarrow\infty$, the case of a single mirror at $z=0$ is recovered. The field modes (\ref{eq:6a}-\ref{eq:6c}) are normalized as

\begin{equation}
\label{eq:6d}
\int \! d^3 r \, \bfm_{\bk j}(\br ) \cdot \bfm_{\bk ' j'}(\br ) = V \delta_{\bk \bk '} \delta_{jj'},
\end{equation}
where $V$ is the quantization volume.

The polarization sum can be conveniently done exploiting the following relation \cite{power82}
\begin{eqnarray}
\label{sumpol}
&\ & \int \! d\Omega \sum_j [\bfm_{\bk j}(\br )]_\ell [\bfm_{\bk j}(\br ')]_m
\nonumber \\
&=& \int \! d\Omega \, \Re \Big\{ \left( \delta_{\ell m}-\hat{k}_\ell \hat{k}_m \right) e^{i\bk \cdot (\br -\br ')}
\nonumber \\
&\ & - \sigma_{\ell p} \left( \delta_{p m}-\hat{k}_p \hat{k}_m \right) e^{i\bk \cdot (\br -\sigma \br ')} \Big\} ,
\end{eqnarray}
where
\begin{eqnarray}
\label{eq:10}
\sigma=
\begin{pmatrix}
1&0&0\\
0&1&0\\
0&0&-1
\end{pmatrix} \, ,
\end{eqnarray}
is the reflection matrix on the reflecting plate placed at $z=0$. For the validity of the relation \eqref{sumpol}, it is understood that all other quantities present inside the angular integration are invariant under the transformation $\bk \rightarrow -\bk$.

We now extend the relations above to the time-dependent case of an oscillating plate, under the adiabatic approximations defined in section \ref{sec:2}. We choose the laboratory frame, where the atoms are at rest and the mirror is moving. The mirror is moving along the $z$ direction with amplitude $a$ and angular frequency $\omega_p$, as discussed in section \ref{sec:2}. We indicate with $\bfm_{\bk j}(\br ,t)$ the instantaneous modes of the form (\ref{eq:6a}-\ref{eq:6c}), relative to time $t$ (they change according to the wall's position oscillating around $z=0$), to be used in the electric field operator \eqref{eq:6} for our dynamical case. The relation we use in our calculations in section \ref{sec:2} is \eqref{sumpol}, appropriately generalized to our (adiabatic) dynamical case. This is done by taking into account that in the right-hand-side, the quantity $\br -\sigma \br '$ in the second term is the distance between the point $\br$ and the image of the point $\br '$ reflected on the mirror; this is a time-dependent quantity because the position of the mirror changes with time. Thus we use the following relation
\begin{eqnarray}
\label{eq:9}
&\ & \int \! d\Omega \sum_j [\bfm_{\bk j}(\br_u,\tp )]_\ell[\bfm_{\bk j}(\br_v,\tpp )]_m\nonumber\\
&=& \int \! d\Omega \, \Re \Big\{(\delta_{\ell m}-{\hat \bk_\ell}{\hat \bk_m})e^{i\bk\cdot(\br_u-\br_v)}\nonumber\\
&\ & -\sigma_{\ell p}(\delta_{p m}-{\hat \bk_p}{\hat \bk_m})e^{i\bk\cdot(\br_u(\tp)-\sigma\br_v(\tpp))} \Big\}
\end{eqnarray}
($\ell,m,p=x,y,z$). In (\ref{eq:9}), $\br_{u(v)}$ $(u,v=A,B)$ is the position vector of atom $A$ or $B$,
$\br_{u}(t)=\br_{u}-{\bf a}\sin(\omega_p t)$ is the instantaneous time-dependent atom-wall distance, and we have defined the vector ${\bf a}=(0,0,a)$.

Relations \eqref{sumpol} and \eqref{eq:9} are obtained in the limit $L \rightarrow \infty$, where the case of a single mirror is recovered. For the validity of \eqref{eq:9}, the conditions mentioned after (\ref{sumpol}) should be verified, as well as our adiabatic approximation.

We wish to point out that we are not including a time dependence of the eigenfrequencies $\wk$, as one could in principle expect for a cavity with a time-dependent length, because at the end we take the limit $L \rightarrow \infty$, and in this limit the field eigenfrequencies have a continuous and time-independent spectrum.

\section*{Acknowledgements}
The authors gratefully acknowledge financial support from the Julian Schwinger Foundation and MIUR.

\end{document}